\documentclass[aps,preprint,showkeys,a4paper,tightenlines,superscriptaddress,12pt]{revtex4}
\usepackage{eurosym}
\usepackage{amsfonts}
\usepackage{amsmath}
\usepackage{amssymb}
\usepackage{graphicx}
\usepackage{subfig}
\usepackage{caption}
\usepackage{setspace}

\providecommand{\U}[1]{\protect\rule{.1in}{.1in}}
\newtheorem{theorem}{Theorem}
\newtheorem{acknowledgement}[theorem]{Acknowledgement}

\begin{document}

\title{Modeling the adhesive contact of rough soft media with an advanced asperity model}
\author{G. Violano}
\affiliation{Department of Mechanics, Mathematics and Management, Polytechnic University
of Bari, Via E. Orabona, 4, 70125, Bari, Italy}
\author{L. Afferrante}
\email{luciano.afferrante@poliba.it}
\affiliation{Department of Mechanics, Mathematics and Management, Polytechnic University
of Bari, Via E. Orabona, 4, 70125, Bari, Italy}

\begin{abstract}
Adhesive interactions strongly characterize the contact mechanics of soft
bodies as they lead to large elastic deformations and contact instabilities.

In this paper, we extend the Interacting and Coalescing Hertzian Asperities
(ICHA) model to the case of adhesive contact. Adhesion is modeled according
to an improved version of the Johnson, Kendall \& Roberts (JKR) theory, in
which jump-in contact instabilities are conveniently considered as well as
the lateral interaction of the asperities and the coalescence of merging
contact spots.

Results obtained on complex fractal geometries with several length scales
are accurate as demonstrated by the comparison with fully numerical
simulations and experimental investigations taken from the literature. Also,
the model quite well captures the distributions of the contact stresses,
gaps, and contact spots.
\end{abstract}

\keywords{Adhesion; rough contact mechanics; JKR theory, fractal surfaces.}
\maketitle

\section{Introduction}

Modeling adhesion between elastic media with rough surfaces is a demanding
challenge. Bodies surfaces may be rough on several length scales, with
roughness amplitudes ranging from nano to micrometric scale. Adhesion is of
central importance in the design of high-technology devices, such us micro
electromechanical systems (MEMS) \cite{mems}, dry adhesives \cite{dry},
stretchable electronics \cite{stretchable}, biomimetic devices \cite%
{biomimetic}. Moreover, adhesion plays a crucial role in the field of
biomaterials, which are developed for drug delivery, medical diagnostics and
tissue engineering \cite{biomaterials}. More in general, tribological
phenomena, e.g. friction and leak-rate of seals, are influenced by the
presence of interfacial adhesion forces \cite{persson}.

In this work, we focus on short-range type adhesion that is typical of soft
elastic materials with high surface energy, where adhesion may lead to
significative elastic deformations of the contacting bodies. In this range,
the Johnson, Kendall \& Roberts (JKR) theory \cite{JKR}, which assumes
infinitely short-range adhesion, is widely believed to be valid.

Numerous authors developed theoretical and numerical approaches to study the
adhesive contact mechanics of rough surfaces. In the framework of the
JKR-type adhesion, we mention the work of Fuller \& Tabor (FT) \cite{FT},
who extended the Greenwood \& Williamson (GW) multiasperity model \cite{GW}
to the adhesive case. More recently, great progresses have been obtained
with the multiscale theory of Persson \cite{persson2002, perstosatti}, where
the detachment force is assumed proportional to an effective interfacial
energy, and in brute-force numerical approaches \cite{CMC17}, like those
developed by Muser \cite{CampanaMuser, ProdanovDappMuser, muser}, where he
shows that "\emph{short-range adhesion compactifies contact patches, changes
various microscopic distribution functions and enhances energy dissipation}%
". Moreover, Rey et al. \cite{rey} with a FFT-based BEM\ methodology
demonstrated that the coefficient of proportionality of the area-load
relation increases with the surface energy.

In this work, adhesion is introduced in the Interacting and Coalescing
Hertzian Asperities (ICHA) model \cite{ICHA1, affpersson}, according to the
JKR theory, as modified in Ref. \cite{CBG} to take into account jump into
contact instabilities occurring when the local gap falls below a critical
value. Experimental and numerical investigations showed that contact jumps
are predominant in the adhesion of very soft bodies \cite{wu, Greensphere}
and are typically due to long-range adhesion interactions originally
neglected in the JKR paper.

\section{The model}

The fundamental equations of the JKR\ theory \cite{JKR}, which relate the
contact load $F$, the contact radius $a$, and the contact approach $\delta $%
, are%
\begin{eqnarray}
F &=&\frac{4}{3}\frac{E^{\ast }a^{3}}{R}-\sqrt{8\pi E^{\ast }\Delta \gamma
a^{3}}  \label{Fjkr} \\
\delta &=&\frac{a^{2}}{R}-\sqrt{\frac{2\pi a\Delta \gamma }{E^{\ast }}}
\label{djkr}
\end{eqnarray}%
where $R$ is the effective radius of curvature, $E^{\ast }$ is the composite
elastic modulus of the contacting bodies and $\Delta \gamma $ is the
interface adhesion energy.

The classical JKR model predicts the first contact occurs when the approach
is zero and, hence, it neglects jump into contact instabilities. Wu \cite{wu}
investigated the jump-in instability occurring in atomic force microscopy
measurements, founding that jump-in is reached at a critical gap $\Delta
_{IN}$. He proposed an empirical formula for the jump-in distance (valid for 
$\mu \geq 2$)%
\begin{equation}
\Delta _{IN}=\left( 1-2.641\mu ^{3/7}\right) \epsilon  \label{dON}
\end{equation}%
where $\mu =\left( \Delta \gamma ^{2}R/E^{\ast ^{2}}\right) ^{1/3}/\epsilon $
is the so-called Tabor parameter \cite{tabor} and $\epsilon $ is the range
of attractive forces.

Ciavarella et al. \cite{CBG}, on the base of the Wu's findings, suggested to
add the effect of van der Waals interactions in the JKR theory by using
equation \ref{dON} for the jump-in critical distance.

\begin{figure}[tbp]
\begin{center}
\includegraphics[scale=0.7]{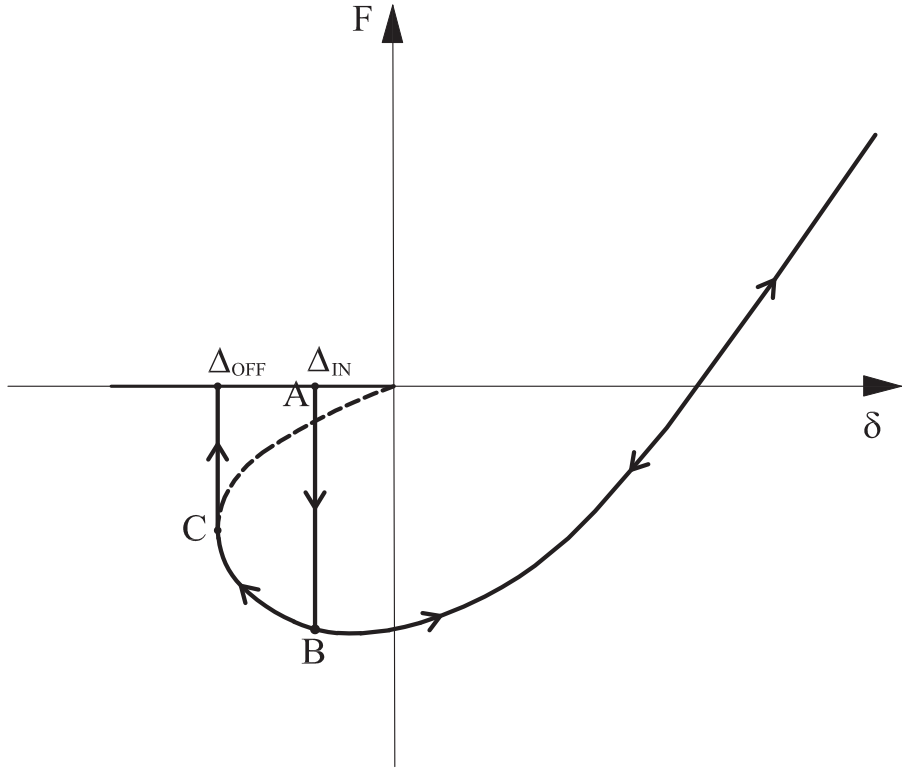}
\end{center}
\caption{The force-displacement relation as predicted by the JKR theory. The
jump-in (point A) and jump-off (point C) are also showed.}
\label{fig1}
\end{figure}

Figure \ref{fig1} shows how the JKR force-displacement relation modifies
introducing the jump-in instability. During the loading phase, when $\delta $
approaches the jump-in value $\Delta _{IN}$, snap-to-contact occurs at the
point \textrm{A} and a sharp decrease in the contact force is observed
(negative $F$ corresponds to tensile force). During the unloading phase,
when the approach $\delta $ reduces up to the jump-off value $\Delta _{OFF}$
(point \textrm{C}) unstable detachment occurs and the contact force and area
vanish.

For randomly rough surfaces, multiple unstable jumps into contact are
usually observed during the loading phase. Therefore, a multiasperity theory
aimed at investigating the adhesive contact mechanics of soft media should
take account of such phenomenon.

The \textit{Interacting and Coalescing Hertzian Asperities} (ICHA) model,
presented for the first time in Ref. \cite{ICHA1}, is an advanced
multiasperity based model, which showed to be quite accurate and efficient
in predicting the main contact quantities of adhesiveless \cite%
{CMC17,affpersson} and DMT-type adhesive \cite{viol1,viol2,viol3} rough
contacts.

In presence of short-range adhesive interactions, DMT-type approaches fail.
For this reason, here, we propose to introduce adhesion in the ICHA model
according to the JKR theory, conveniently modified to also consider jump-in
instabilities, as suggested in Ref. \cite{CBG}.

Consider a rigid rough media approaching to an elastic half-space. Following
the JKR formalism, the normal displacement $w_{i}$ of the elastic half-space
at the location of the asperity $i$ can be written as%
\begin{equation}
w_{i}=\frac{a_{i}^{2}}{R_{i}}-\sqrt{\frac{2\pi a_{i}\Delta \gamma }{E^{\ast }%
}}+\hat{w}_{i}  \label{wICHA}
\end{equation}%
where%
\begin{eqnarray}
\hat{w}_{i} &=&\sum_{j=1,j\neq i}^{n_{ac}}\frac{a_{j}^{2}}{\pi R_{j}}\left( 
\sqrt{\frac{r_{ij}^{2}}{a_{j}^{2}}-1}+\left( 2-\frac{r_{ij}^{2}}{a_{j}^{2}}%
\right) \arcsin \left( \frac{a_{j}}{r_{ij}}\right) \right)  \notag \\
&&-\frac{1}{\pi a_{j}E^{\ast }}\sqrt{8\pi a_{j}^{3}E^{\ast }\Delta \gamma }%
\arcsin \left( \frac{a_{j}}{r_{ij}}\right)  \label{wint}
\end{eqnarray}%
is the displacement due to the elastic interaction between all the
asperities in contact $n_{ac}$. In eq. \ref{wint}, $r_{ij}$ is the distance
between the asperities $i$ and $j$.

The jump-in distance given in eq. \ref{dON} is calculated for each asperity.
When the rough surface approaches the half-space, contact occurs when the
gap between an asperity and the half-space becomes smaller than $\Delta
_{IN} $. The first estimate of the asperity contact radius is done by
inverting the JKR relation \ref{djkr}. Then, after a further increment of
the approach $\eta _{i}$, the contact radius is increased by the quantity%
\begin{equation}
\Delta a_{i}=\frac{\eta _{i}}{2a_{i}/R_{i}-\sqrt{\pi \Delta \gamma /(E^{\ast
}a_{i})}}  \label{da}
\end{equation}%
which is obtained by differentiating eq. \ref{wICHA}.

The coalescence of merging contact patches is taken into account as
described in Ref. \cite{ICHA1}. Asperities with overlapping contact spots
are replaced with a new equivalent one, which maintains the same total
contact area of the suppressing asperities ($a_{eq}^{2}=a_{i}^{2}+a_{j}^{2}$%
). Moreover, the position of the volume centroid is kept unchanged and the
equivalent radius of curvature $R_{eq}$ is empirically assumed as $\left(
R_{i}^{2}+R_{j}^{2}\right) ^{1/2}$. Finally, the height $h_{eq}$ of the new
asperity is defined so that the contact area is effectively $\pi a_{eq}^{2}$
at the given separation.

The total contact area and the total load are then obtained by summing up
the contributions of all the asperities in contact.

Moreover, since a self-balanced load distribution is considered, the
interfacial mean separation $\bar{u}$ is computed as $\bar{u}_{0}-\delta $,
where $\bar{u}_{0}$ and $\delta $ are the initial separation and the total
approach, respectively.

The local displacement at the location of a point $Q$ outside the contact
region can be calculated as%
\begin{eqnarray}
w(Q) &=&\sum_{k=1}^{n_{ac}}\frac{a_{k}^{2}}{\pi R_{k}}\left( \sqrt{\frac{%
r_{Qk}^{2}}{a_{k}^{2}}-1}+\left( 2-\frac{r_{Qk}^{2}}{a_{k}^{2}}\right)
\arcsin \left( \frac{a_{k}}{r_{Qk}}\right) \right)  \notag \\
&&-\frac{1}{\pi a_{k}E^{\ast }}\sqrt{8\pi a_{k}^{3}E^{\ast }\Delta \gamma }%
\arcsin \left( \frac{a_{k}}{r_{Qk}}\right)  \label{wQ}
\end{eqnarray}%
where $r_{Qk}$ is the distance between the asperity $k$ and the point $Q$.

Moreover, as in the JKR theory adhesive stresses are supposed acting only
inside the contact area $\pi a^{2}$, the contact pressure is obtained by
superposing the Hertzian repulsive contribution and that due to a flat rigid
cylindrical punch of the same radius $a$. Therefore, at distance $r$ from
the centre of the contact spot, the contact pressure is%
\begin{equation}
p(r)=2\frac{E^{\ast }a}{\pi R}\left( 1-\frac{r^{2}}{a^{2}}\right)
^{1/2}-\left( 2\frac{E^{\ast }\Delta \gamma }{\pi a}\right) ^{1/2}\left( 1-%
\frac{r^{2}}{a^{2}}\right) ^{-1/2}\text{ \ \ for }r\leq a.  \label{prJKR}
\end{equation}

\section{Adhesion of self-affine fractal surfaces}

The power spectral density (PSD) of natural surfaces often follows a
self-affine behavior \cite{fractal}. For this reason, we have performed
calculations on nominally flat surfaces with roughness\ described by a
self-affine fractal geometry with PSD%
\begin{eqnarray}
C(q) &=&C_{0}\qquad \qquad \qquad \qquad \text{for }q_{L}\leq q<q_{0}  \notag
\\
C(q) &=&C_{0}\left( q/q_{0}\right) ^{-2(H+1)}\qquad \text{for }q_{0}\leq
q<q_{1}  \label{PSD}
\end{eqnarray}

Surfaces have been generated according to the spectral methodology developed
in Ref. \cite{put1, put2}.

In eq. \ref{PSD}, $q$ is the modulus of the wave vector $\mathbf{q}%
=(q_{x},q_{y})$, $q_{L}=2\pi /L$ (being $L$ the lateral size of the domain)
and $q_{1}$ are the short and long frequencies cut-off, respectively, while $%
q_{0}$ is the roll-off frequency. The exponent $H$ is the \textit{Hurst} 
\textit{exponent}, which is related to the fractal dimension by $D_{f}=3-H$.

\subsection{Comparison with the Contact Mechanics Challenge's data}

The Contact Mechanics Challenge (CMC) \cite{CMC17}, proposed by M\"{u}ser to
the tribology community in \textrm{2015}, consisted in simulating the
adhesive contact between a rigid self-affine fractal surface and an elastic
half-space. Several scientists participated submitting data obtained by
their numerical, analytical and experimental methodologies.

One of the authors of this paper contributed with data obtained by the ICHA\
model; however, in that circumstance, the ICHA\ simulations were performed
neglecting adhesion.

Since the parameters adopted in the CMC were chosen to mimic the short-range
adhesion contact between a rubber and a polished surface, the conditions are
close to the JKR limit. For this reason, one can expect that a JKR-like
adhesion model is appropriate for studying the problem. Therefore, here we
compare the results predicted by the proposed model with the reference data
obtained with the Green Function Molecular Dynamics (GFMD) code by M\"{u}ser 
\cite{CampanaMuser, ProdanovDappMuser}.

\begin{figure}[tbp]
\begin{center}
\includegraphics[width=15.0cm]{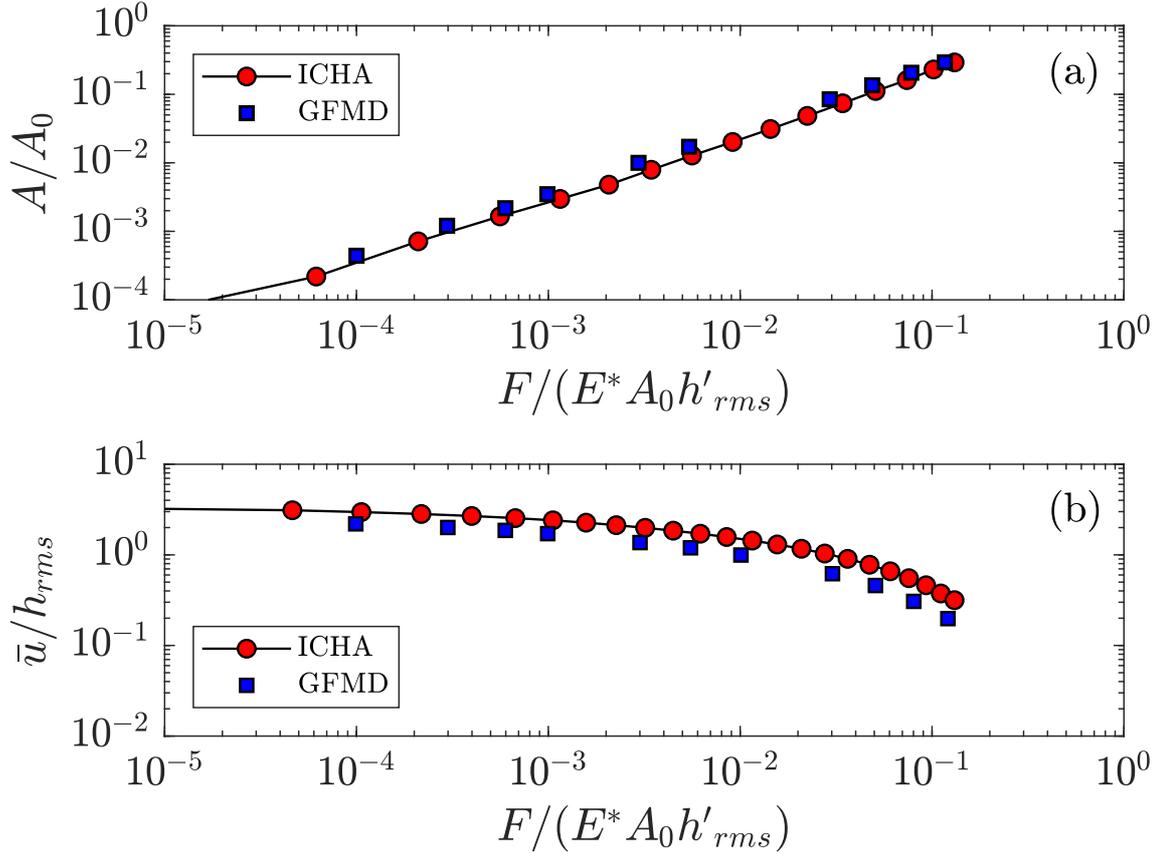}
\end{center}
\caption{(a): The normalized real contact area $A/A_{0}$ as a function of
the dimensionless load $F/(E^{\ast }h_{\mathrm{rms}}^{\mathrm{\prime }}A_{0})
$. Results are referred to the ICHA model (red markers) and GFMD\ code (blue
markers). (b): The normalized mean interfacial separation $\bar{u}/h_{%
\mathrm{rms}}$ as a function of the dimensionless load $F/(E^{\ast }h_{%
\mathrm{rms}}^{\mathrm{\prime }}A_{0})$. Results are referred to the ICHA
model (red markers) and GFMD\ code (blue markers).}
\label{fig2}
\end{figure}

Fig. \ref{fig2}a shows, in a double-logarithmic plot, the normalized real
contact area $A/A_{0}$ as a function of the dimensionless pressure $%
F/(E^{\ast }h_{\mathrm{rms}}^{\mathrm{\prime }}A_{0})$, being $A_{0}$ the
nominal contact area and $h_{\mathrm{rms}}^{\mathrm{\prime }}$ the root mean
square (rms) slope of the surface. Results show linearity in a large range
of loads and the agreement with GFMD simulations is quite satisfactory.

In Fig. \ref{fig2}b the interfacial mean separation $\bar{u}$, normalized
with respect to the rms roughness amplitude $h_{\mathrm{rms}}$, is plotted
as a function of dimensionless pressure $F/(E^{\ast }h_{\mathrm{rms}}^{%
\mathrm{\prime }}A_{0})$, in a double-logarithmic representation.
Predictions of $\bar{u}/h_{\mathrm{rms}}$ with the ICHA\ model slightly
exceed the reference solution, but once again the general trend is in good
agreement.

\begin{figure}[tbp]
\centering
\includegraphics[width=15.0cm]{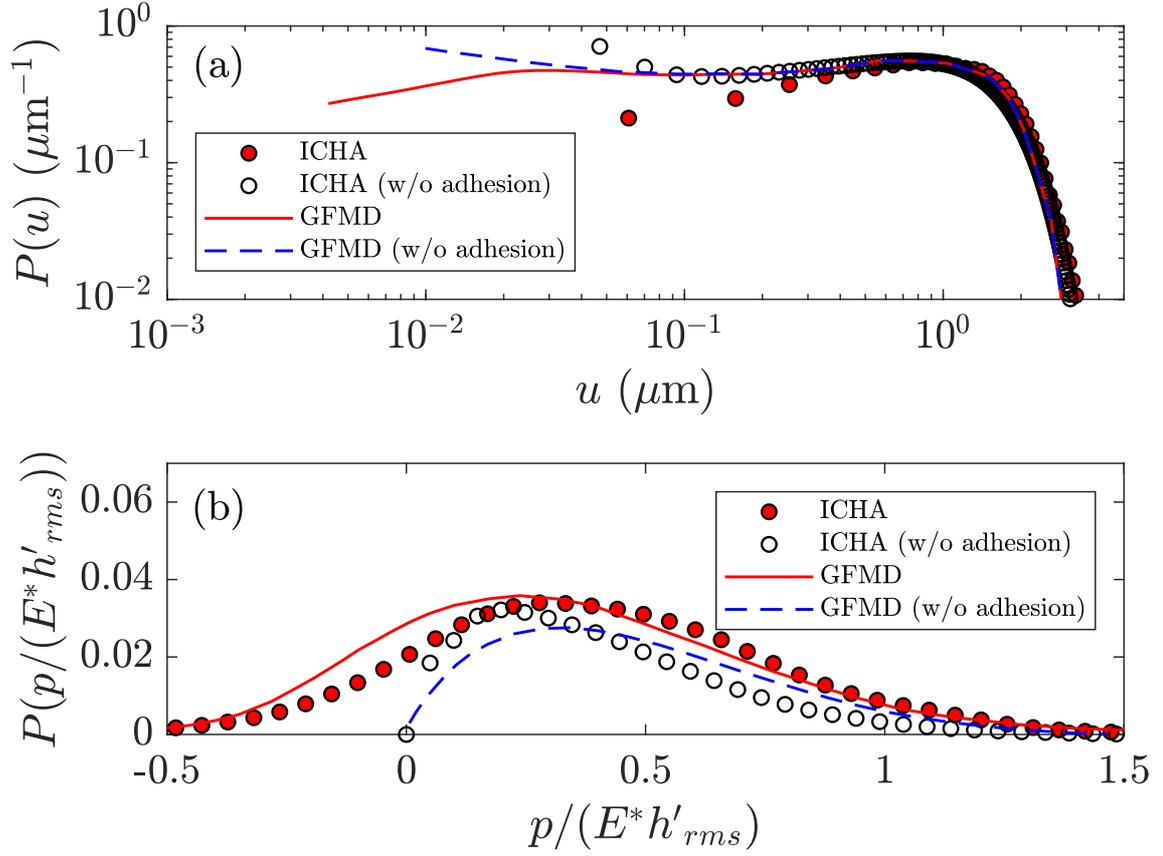}
\caption{(a): The gap probability distribution $P(u)$. Results are referred
to the ICHA model with adhesion (red circular markers) and without adhesion
(white circular markers), GFMD\ code without adhesion (blue dashed line) and
GFMD\ code with adhesion (red solid line). (b): The probability distribution
of the dimensionless interfacial pressures $P(p/(E^{\ast }h_{\mathrm{rms}}^{%
\mathrm{\prime }}))$. Symbol details are the same of (a). Both in (a) and
(b) results are obtained at a load $F/(E^{\ast }h_{\mathrm{rms}}^{\mathrm{%
\prime }}A_{0})=0.01$.}
\label{fig3}
\end{figure}

Fig. \ref{fig3}a shows the gap probability distribution $P(u)$ in a
double-logarithmic chart. Results agree with the GFMD\ data with and without
adhesion. Some discrepancies are observed only at low gaps. However, we
observe that the accuracy of predictions is strongly influenced by the
discretization of the surface mesh. In this respect, the solution shows the
typical behavior of a short-range adhesion, which leads to a strongly
reduced probability for small gaps. The latter is due to the formation of
JKR adhesive necks near the contact line. To capture the formation of such
necks, the calculations of the GFMD\ code were performed on systems with $%
16\times 10^{9}$ discretization points on the surface. The present
simulations were instead carried out using only$\ 512\times 512=262144$ grid
points.

Fig. \ref{fig3}b shows the probability distribution of the dimensionless
interfacial pressures $P(p/(E^{\ast }h_{\mathrm{rms}}^{\mathrm{\prime }}))$.
The lower number of discretization points could in part justify the
differences observed at negative values of the pressure, where the ICHA\
model underestimates $P(p/(E^{\ast }h_{\mathrm{rms}}^{\mathrm{\prime }}))$.
Indeed, negative values of the pressure are expected at the edge of contact
areas, i.e., at the location of JKR adhesive necks. Notice the pronounced
peak at small negative pressures predicted in Ref. \cite{CMC17} and
reflecting that "\textit{most non-contact points have an interfacial
separation that greatly exceeds the range of the adhesive interaction}", is
absent in our simulations. In fact, the contribution to $P\left( p/(E^{\ast
}h_{\mathrm{rms}}^{\mathrm{\prime }})\right) $ is originated only from
contact regions as in our model stresses vanish in the non-contact points.
For this reason, the comparison is proposed with the probability distribution
originating from the true contact area.

\begin{figure}[tbp]
\centering
\includegraphics[width=10.0cm]{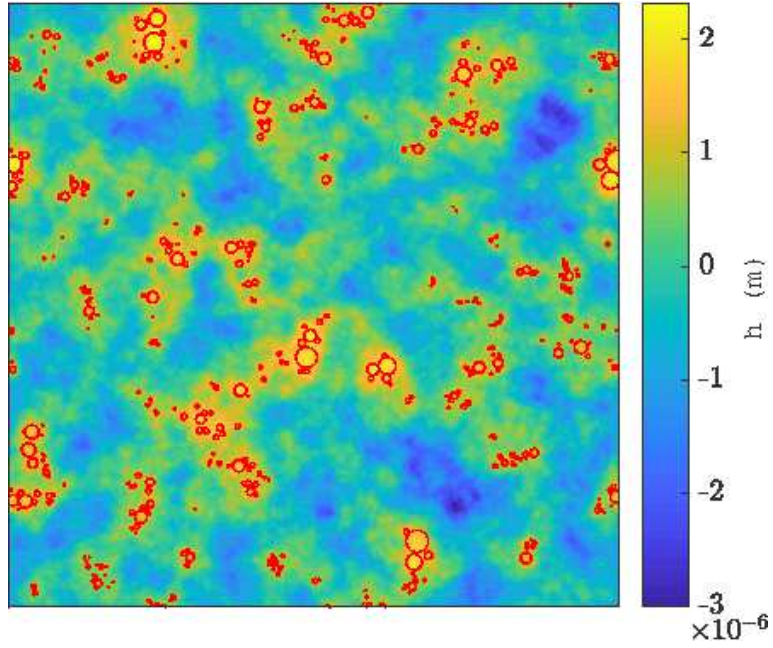}
\caption{The superposition of the contact spots predicted by the ICHA\ model
(red circles) on the surface
topography. Results are obtained at a load $F/(E^{\ast }h_{\mathrm{rms}%
}^{\mathrm{\prime }}A_{0})=0.01$.}
\label{fig4}
\end{figure}

Fig. \ref{fig4} shows the contact spots predicted by the ICHA\ model (red circles) accurately follow the surface topography as they are located on the highest peaks of the surface (yellow regions).

\subsection{Comparison with the experimental measurements of McGhee et al.}

McGhee et al. \cite{McGhee} measured the area-load relation in the contact
between rough surfaces and smooth elastic PDMS\ samples. The CMC surface was
produced on \textrm{1000} times scaled models with three different rms
slopes $h_{\mathrm{rms}}^{\mathrm{\prime }}=$ $0.2$, $0.5$ and $1$. Optical
measurements of the contact area were performed using frustrated total
internal reflectance. Moreover, different PDMS\ samples with various values
of the composite elastic modulus $E^{\ast }$ and surface energy $\Delta
\gamma $ were manufactured.

\begin{figure}[tbp]
\centering
\includegraphics[width=13.0cm]{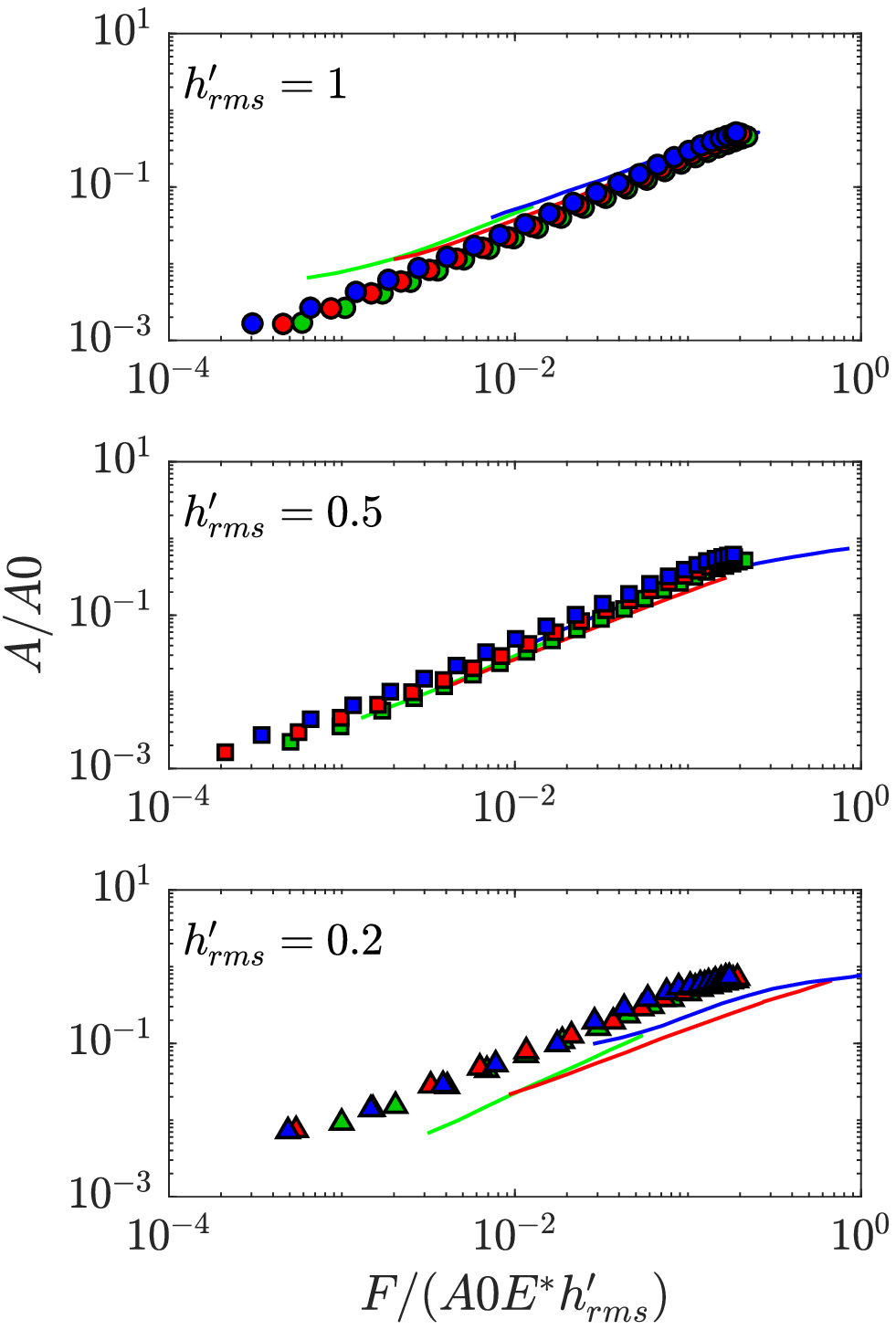}
\caption{The normalized real contact area $A/A_{0}$ as a function of the
dimensionless load $F/(E^{\ast }h_{\mathrm{rms}}^{\mathrm{\prime }}A_{0})$.
We compare the ICHA model\ predictions (markers) and the experimental
measurements of McGhee et al. (solid lines). Results are shown for three
values of the rms slope $h_{\mathrm{rms}}^{\mathrm{\prime }}=$ $1$
(circles), $h_{\mathrm{rms}}^{\mathrm{\prime }}=$ $0.5$ (squares) and $h_{%
\mathrm{rms}}^{\mathrm{\prime }}=$ $0.2$ (triangles). Numerical and
experimental investigations have been performed for different values of $%
E^{\ast }$ and $\Delta \protect\gamma $; in particular we have used $E^{\ast
}=0.24$ $\mathrm{MPa}$, $\Delta \protect\gamma =2$ \textrm{mJ/m}$^{2}$ (blue
data), $E^{\ast }=0.75$ $\mathrm{MPa}$, $\Delta \protect\gamma =3$ \textrm{%
mJ/m}$^{2}$ (red data) and $E^{\ast }=2.10$ $\mathrm{MPa}$, $\Delta \protect%
\gamma =4$ \textrm{mJ/m}$^{2}$ (green data).}
\label{fig5}
\end{figure}

Fig. \ref{fig5} shows a double-logarithmic representation of the normalized
real contact area $A/A_{0}$ as a function of the dimensionless pressure $%
F/(E^{\ast }h_{\mathrm{rms}}^{\mathrm{\prime }}A_{0})$. Results are
presented for three values of $E^{\ast }$ and $\Delta \gamma $, and at
different values of the rms slope. The lines denote the experimental data
taken from Ref. \cite{McGhee}, while the markers correspond to the
prediction of the ICHA\ model. A good agreement is observed for $h_{\mathrm{%
rms}}^{\mathrm{\prime }}=$ $0.5$ and $1$. Some discordance occurs at $%
h_{\mathrm{rms}}^{\mathrm{\prime }}=$ $0.2$, but we stress that for this
value of the rms slope scattering was found in the experiments.
Moreover, the manufactured surfaces do not truly satisfy the small-slope
approximation as they were produced with the assigned rms gradient of one.
Another potential \textquotedblleft limitation\textquotedblright\ is the
long-time viscoelastic response that can occur in the experiments.

\begin{figure}[tbp]
\begin{center}
\includegraphics[width=16.0cm]{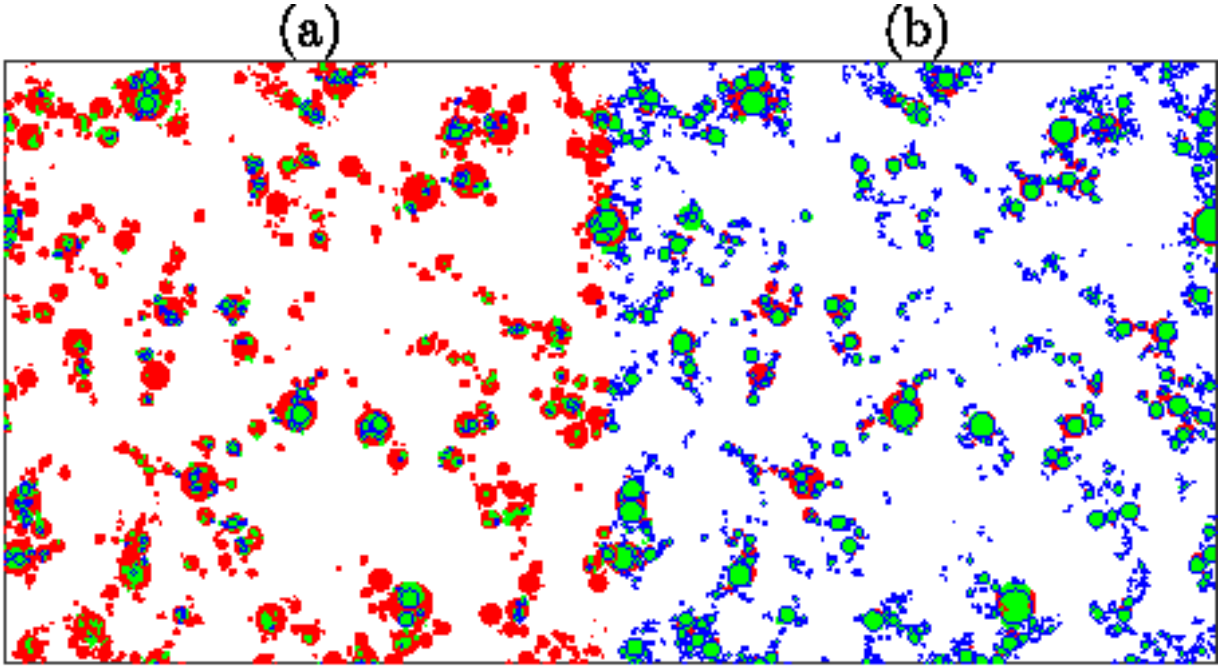}
\end{center}
\caption{The contact spots predicted by the ICHA\ model at fixed load (a)
and fixed contact area (b) for three values of the rms slope $h_{\mathrm{rms}%
}^{\mathrm{\prime }}=$ $1$ (blue spots), $0.5$ (green spots) and $0.2$ (red
spots).}
\label{fig6}
\end{figure}

Increasing the rms slope, a considerable decrease in the contact area is
detected for a fixed value of the applied load. However, variations of the
rms slope do not alter the shape of the contact patches under the condition
of nearly equal contact area. A qualitative explanation of the former
effects is proposed in Fig. \ref{fig6}, where the contact patches predicted
by the ICHA\ model are shown for all the investigated values of $h_{\mathrm{%
rms}}^{\mathrm{\prime }}$. In particular, results are shown at fixed
dimensionless applied load $F/(E^{\ast }A_{0})$ (Fig. \ref{fig6}a) and
normalized contact area $A/A_{0}$ (Fig. \ref{fig6}b).

\section{Conclusions}

We have proposed an extension of the ICHA contact model to the case of
short-range adhesion (JKR limit). In previous works, we have found such
model to be accurate when adhesion is neglected or in the limit of contact
with long-range adhesion (DMT limit). Here, the model confirms accuracy in
predicting the main contact quantities as we infer from the comparison with
data of the contact mechanics challenge \cite{CMC17} and experimental
measurements \cite{McGhee}.

Results show once again the fundamental role played by the elastic coupling
and coalescence of merging contact spots as crucial factors to obtain good
results with asperity models.

Hence, the ICHA model is fast as well as accurate in modeling adhesive
contact problems; however, some limitations of the methodology are also
evident as the adhesion of soft media inherently needs hard computational
and time-consuming efforts for a better representation of the contact.

\begin{acknowledgement}
The authors acknowledge support from the Italian Ministry of Education,
University and Research (MIUR) under the program \textquotedblleft
Departments of Excellence" (L.232/2016).
\end{acknowledgement}

\end{document}